\documentclass[12pt,xcolor]{article}

\usepackage{amssymb,amsmath,amsthm,bm}
\usepackage{url}
\usepackage[numbers]{natbib}
\usepackage{graphicx}
\usepackage[mathscr]{eucal}
\usepackage[OT2,T1]{fontenc}
\usepackage{verbatim}

\newcommand{\mygreen}{\color{green!50!black}}
\newcommand{\myblue}{\color{blue!50!black}}
\newcommand{\myred}{\color{red!50!black}}

\newcommand{\argmin}{\ensuremath{\arg\min}}

\newcommand{\1}{\ensuremath{\mbox{{\bf 1}}}}

\newcommand{\mC}{\mathcal{C}}

\newcommand{\mX}{\mathcal{X}}

\newcommand{\mF}{\mathcal{F}}
\newcommand{\mP}{\mathcal{P}}

\newcommand{\Real}{\mathbb{R}}
\newcommand{\iid}{\stackrel{iid}{\sim}}

\begin{document}

\title{Manifold Matching:\\Joint Optimization of Fidelity and Commensurability}

\author{
Carey E.\ Priebe\footnote{
Corresponding Author: Department of Applied Mathematics and Statistics,
Johns Hopkins University, Baltimore, MD 21218-2682 ; \mbox{ cep@jhu.edu }.
This work is partially supported by
National Security Science and Engineering Faculty Fellowship (NSSEFF),
Air Force Office of Scientific Research (AFOSR),
Office of Naval Research (ONR),
Johns Hopkins University Human Language Technology Center of Excellence (JHU HLT COE),
and the American Society for Engineering Education (ASEE) Sabbatical Leave Program.
}, Johns Hopkins University\\
David J.\ Marchette, Naval Surface Warfare Center\\
Zhiliang Ma, Johns Hopkins University\\
Sancar Adali, Johns Hopkins University
}

\maketitle

\abstract{
Fusion and inference from multiple and massive disparate data sources -- the
requirement for our most challenging data analysis problems and the goal of our most
ambitious statistical pattern recognition methodologies -- has many and varied
aspects which are currently the target of intense research and development. One aspect of the
overall challenge is manifold matching -- identifying embeddings of multiple
disparate data spaces into the same low-dimensional space where joint inference can
be pursued. We investigate this manifold matching task from the perspective of
jointly optimizing the fidelity of the embeddings and their commensurability with
one another, with a specific statistical inference exploitation task in mind. Our
results demonstrate when and why our joint optimization methodology is superior to
either version of separate optimization. The methodology is illustrated with
simulations and an application in document matching.
}

\section{Introduction}

\subsection{Motivation}

Let $(\Xi,\mF,\mP)$ be a probability space,
i.e., $\Xi$ is a sample space, $\mF$ is a sigma-field,
and $\mP$ is a probability measure.
Consider $K$ measurable spaces $\Xi_1,\cdots,\Xi_K$ 
and measurable maps $\pi_k:\Xi \to \Xi_k$.
Each $\pi_k$ induces a probability measure $\mP_k$ on $\Xi_k$.
We wish to identify a measurable metric space $\mX$
(with distance function $d$)
and measurable maps $\rho_k: \Xi_k \to \mX$,
inducing probability measures $\widetilde{\mP}_k$ on $\mX$,
so that for $[x_1,\cdots,x_K]' \in \Xi_1 \times \cdots \times \Xi_K$
we may evaluate distances $d(\rho_{k_1}(x_{k_1}),\rho_{k_2}(x_{k_2}))$ in $\mX$.
See Figure 1.

Given $\xi_1,\xi_2 \iid \mP$ in $\Xi$,
we may reasonably hope that the random variable
$d(\rho_{k_1}\circ\pi_{k_1}(\xi_1),\rho_{k_2}\circ\pi_{k_2}(\xi_1))$
is stochastically smaller than the random variable
$d(\rho_{k_1}\circ\pi_{k_1}(\xi_1),\rho_{k_2}\circ\pi_{k_2}(\xi_2))$.
That is, matched measurements 
$\pi_{k_1}(\xi_1),\pi_{k_2}(\xi_1)$
representing a single point $\xi_1$ in $\Xi$
are mapped closer to each other than are
unmatched measurements 
$\pi_{k_1}(\xi_1),\pi_{k_2}(\xi_2)$
representing two different points in $\Xi$.
This property allows inference to proceed in the common representation space $\mX$.

However, 
we do not observe $\xi \in \Xi$;
we also do not observe the $x_k = \pi_k(\xi) \in \Xi_k$ directly,
nor do we have knowledge of the maps $\pi_k$.
But suppose we have access to functions 
$\delta_k:\Xi_k \times \Xi_k \to \mathbb{R}_+ = [0,\infty)$
such that $\delta_k( \pi_k(\xi_1) , \pi_k(\xi_2) )$
represents the ``dissimilarity'' of outcomes $\xi_1$ and $\xi_2$
under map $\pi_k$.
We propose to use sample dissimilarities for matched data in the disparate spaces $\Xi_k$
to simultaneously learn maps $\rho_k$ which allow for a powerful test of matchedness
in the common representation space $\mX$.

\begin{figure}[h]
  \begin{center}
    \includegraphics[page=1, scale=4.25]{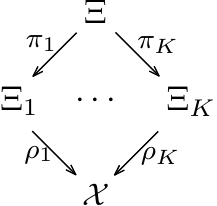}
    \caption{Maps $\pi_k$ induce disparate data spaces $\Xi_k$ from ``object space'' $\Xi$.
    Manifold matching involves using matched data $\{\bm{x}_{ik}\}$
    to simultaneously learn maps $\rho_1,\ldots,\rho_K$
    from disparate spaces 
    $\Xi_1,\ldots,\Xi_K$
	to a common ``representation space'' $\mX$, for subsequent inference.}\label{fig:mm}
  \end{center}
  \end{figure}

\subsection{Problem Formulation}

Consider $n$ objects each measured under $K$ different conditions,
$$\bm{x}_{i1} \sim \cdots \sim \bm{x}_{ik} \sim \cdots \sim\bm{x}_{iK},\ i = 1, \ldots, n,$$
where $\bm{x}_{i1} \sim \cdots \sim \bm{x}_{ik} \sim \cdots \sim \bm{x}_{iK}$
denotes $K$ matched measurements 
$\pi_1(\xi_i),\cdots,\pi_K(\xi_i)$
representing a single object $\xi_i \in \Xi$,
where $\Xi$ denotes the ``object space''.
The assumption of $K$ {\em different} conditions implies that
$\bm{x}_{ik} \in \Xi_k$
where the spaces $\Xi_1,\cdots,\Xi_K$ cannot be assumed to be similar.
We are given $K$ new measurements $\{\bm{y}_k\}_{k=1}^K,\ \bm{y}_k \in \Xi_k$.
The question under consideration is:
Does the collection $\{\bm{y}_k\}_{k=1}^K$ also correspond to matched measurements representing a single object measured under the $K$ conditions?

We use the $\Xi$ notation to remind the reader that the spaces $\Xi_k$ cannot be assumed to be standard finite-dimensional Euclidean spaces.
We do assume that each space $\Xi_k$ comes with a within-condition dissimilarity $\delta_k$ --
a hollow, symmetric function from $\Xi_k \times \Xi_k$ to $\mathbb{R}_+$ --
through which the matched data $\{\bm{x}_{ik}\}$ yields $n \times n$ dissimilarity matrices $\Delta_k$, $k=1,\cdots,K$.
For new measurements $\{\bm{y}_k\}_{k=1}^K$ we have available for each $k$ the within-condition dissimilarities
$\delta_k(\bm{y}_{k}, \bm{x}_{ik}) , \ i = 1, \ldots, n$.

Remark 1: The $\bm{x}_{ik}$ and $\bm{y}_{k}$ are introduced mainly for symbolic purposes;
the corresponding data may not be available or may be too complex to use directly,
and we proceed from the dissimilarities.

The specific statistical inference exploitation task we consider throughout most of this article is hypothesis testing.
Our goal, simplified for the case $K=2$, is to determine whether $\bm{y}_{1}$ and $\bm{y}_{2}$ are a match.
That is, $$H_0: \bm{y}_{1} \sim \bm{y}_{2} \ \text{ versus } \ H_A: \bm{y}_{1} \nsim \bm{y}_{2},$$
or equivalently,
$$H_0: \bm{y}_{1}=\pi_1(\xi) , \bm{y}_{2}=\pi_2(\xi) \ \text{ versus } \ H_A: \bm{y}_{1}=\pi_1(\xi) , \bm{y}_{2}=\pi_2(\xi') \mbox{ ~ for ~ } \xi \neq \xi' \in \Xi.$$
(We control the probability of missing a true match.)

\subsection{Manifold Matching}

We define {\em manifold matching}
as simultaneous manifold learning and manifold alignment --
identifying embeddings of multiple disparate data sources into the same low-dimensional space
where joint inference can be pursued.
Figure \ref{fig:mm} depicts our framework.
Conditional distributions are induced by maps $\pi_k$ from ``object space'' $\Xi$.
Our assumption is that the conditional spaces $\Xi_k$ are {\em not} commensurate.
For example, if the elements of $\Xi$ are individual people, then a photograph in image space $\Xi_1$
and a biographical sketch in text document space $\Xi_2$ are not to be directly compared.
Indeed, our fundamental premise defining {\em disparate} data sources is that
the various $\Xi_k$ cannot profitably be treated as replicates of the same kind of space.
Rather, the various spaces are different not just in degree but in kind.
Each dissimilarity $\delta_k$ has been tailored for application to $\Xi_k$,
and it is inappropriate to apply $\delta_{k}$ on $\Xi_k \times \Xi_{k'}$ for $k' \neq k$.
This distinguishes our {\em data fusion} from conventional multivariate analysis.

In Figure \ref{fig:mm},
matched points $\{\bm{x}_{ik}\}$ are used to simultaneously learn appropriate maps $\rho_k$
taking the disparate data from the various $\Xi_k$ into a common representation space $\mathcal{X}$.
These maps are then applied to $\{\bm{y}_k\}_{k=1}^K$
yielding $\widetilde{\bm{y}}_k = \rho_k(\bm{y}_k)$,
whence (for $K=2$) we use $T=d(\widetilde{\bm{y}}_1,\widetilde{\bm{y}}_2)$
as our test statistic and reject for $T$ ``large''.

Remark 2: Our convention is to use the ``~$\widetilde{\cdot}$~'' notation for points in the target space $\mathcal{X}$,
contrasted with no tilde for points in the original $\Xi_k$ spaces.


Remark 3: We will throughout consider the special case of $\mathcal{X} = \mathbb{R}^m$
for some pre-specified target dimension $m$.
The fundamentally important and challenging task of choosing the target dimension --
{\em model selection} -- will be considered only as a confounding issue in this paper;
$m$ is a nuisance parameter which must be selected but whose selection is beyond the scope of this manuscript.

\subsection{What are these ``conditions'' and what does ``matched'' mean?}

As suggested above,
one example of ``conditions'' involves
photographs $\{\bm{x}_{i1}\}$
and biographical sketches $\{\bm{x}_{i2}\}$,
with ``matched'' $\bm{x}_{i1} \sim \bm{x}_{i2}$
meaning that the photograph $\bm{x}_{i1}$ and the biographical sketch $\bm{x}_{i2}$
are of the same person.

Other illustrative examples include:
a general image \& caption scenario,
with ``matched'' meaning that they go together;
multiple languages for text documents, with ``matched'' meaning on the same topic;
multiple modalities for photographs
(e.g., indoor lighting vs outdoor lighting,
two cameras of different quality, or
passport photos and airport surveillance photos),
with ``matched'' meaning of the same person;
Wikipedia text document and Wikipedia hyperlink structure,
with ``matched'' meaning of the same document.
More generally, our framework may be applicable to any scenario in which multiple dissimilarity measures are applied to the objects at hand.

\begin{figure}[h]
\begin{center}
\includegraphics[scale=0.85]{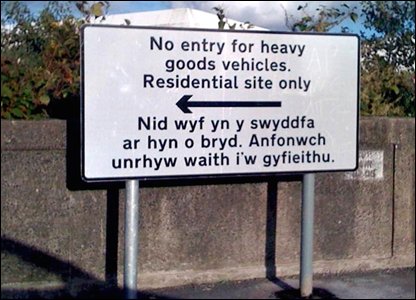}
\caption{
An example of ``not matched'' for multi-lingual text documents.
The English is clear enough to lorry drivers --- but the Welsh reads
``I am not in the office at the moment. Send any work to be translated.''
(See \protect\url{http://news.bbc.co.uk/2/hi/uk_news/wales/7702913.stm};
permission obtained from \protect\url{http://www.golwg360.com/Hafan/default.aspx}.)
}\label{fig:notmatched}
\end{center}
\end{figure}

Fundamentally, ``matched'' means whatever the training data say it means.
We know it when we see it --
or, perhaps more accurately, we know {\em un}matched when we see it; see Figure \ref{fig:notmatched}.
Consider, for instance, an example of
multiple languages for text documents, with ``matched'' meaning on the same topic.
Given English and French Wikipedia documents with the matching provided by Wikipedia itself,
``matched'' means ``on the same topic.''
But of course the Wikipedia documents are not direct translations of one another,
and documents in different languages on the same topic may have significant conceptual differences
due to cultural differences, etc.

\subsection{Dirichlet Setting}\label{section:dirichlet}

While the matched training data ultimately determine what ``matched'' means,
in order to provide a clear mathematical characterization of matchedness
we consider an illustrative Dirichlet setting.
This setting is clearly overly simplified, but it invokes some aspects of
the foregoing example of multiple languages for text documents.

\begin{figure}[h]
    \begin{center}
      \includegraphics[scale=1.2]{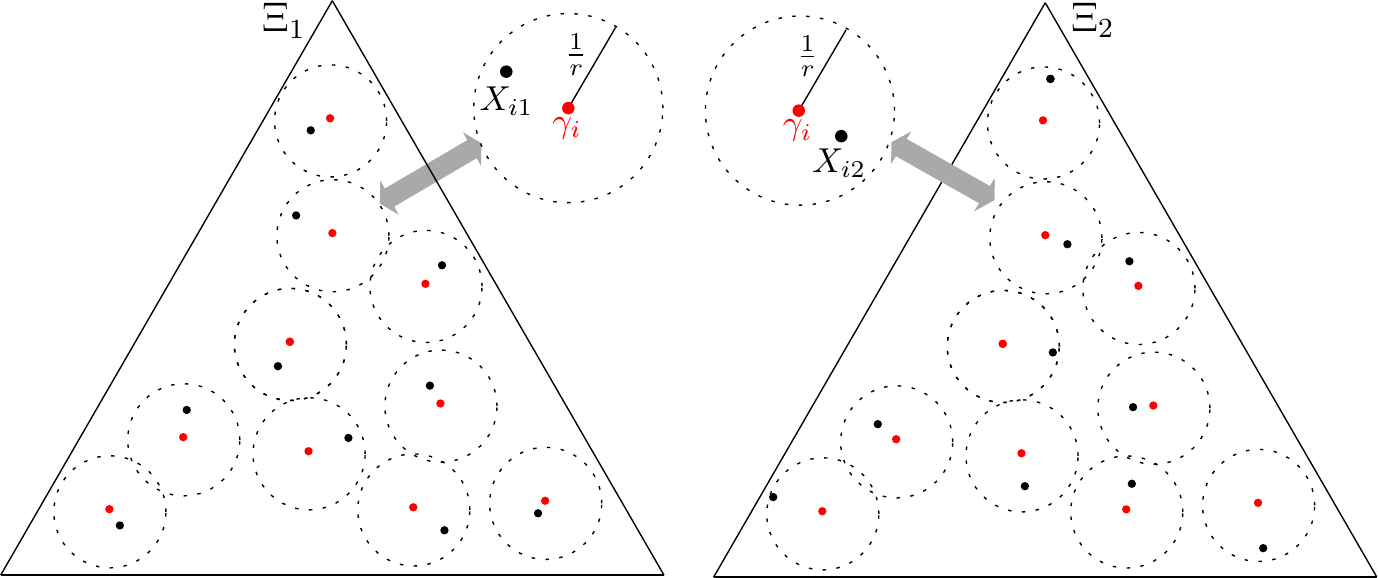}
    \caption{Illustrative Dirichlet setting wherein
    $X_{ik} \iid Dirichlet(r\bm{\gamma}_i+\1)$ represent documents $i=1,\ldots,n=10$ in languages $k=1,\ldots,K=2$ in the standard 2-simplex $S^2$.
  The parameter $r$ controls the meaning of matchedness --
the similarity of matched documents $X_{i1}$ and $X_{i2}$
compared to unmatched documents $X_{i1}$ and $X_{i'2}$ for $i \neq i'$.
    }\label{fig:dirichlet}
  \end{center}
\end{figure}

\begin{figure}[h]
\begin{center}
\includegraphics[scale=0.72]{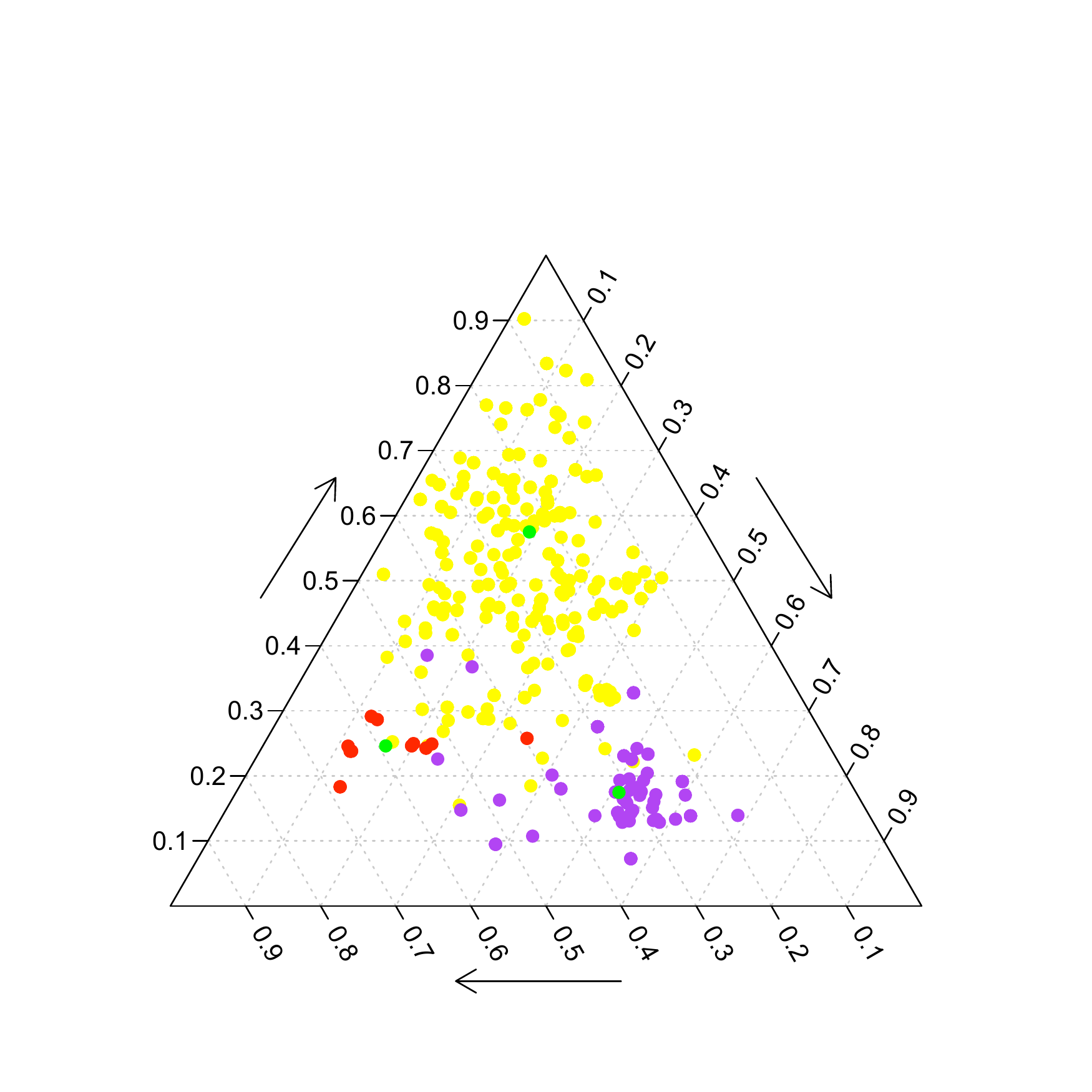}
\caption{An example considering the FLD$\circ$LSA projection into $S^2$ of multiple
Enron email messages identified with three Linguistic Data Consortium (LDC) topics.
The three colored scatterplots -- yellow, red, purple -- represent documents from the three topics;
the green dots represent the topic means.
We see that ``matched'', meaning ``on the same topic'', does mean something quite like $Dirichlet(r\bm{\gamma}_{topic}+\1)$ in this case
(but the variability ``$r$'' may be topic-dependent).
}\label{fig:MingSun1}
\end{center}
\end{figure}

  Let
  $S^p =\{\bm{x} \in \mathbb{R}^{p+1}_+: \sum_{\ell=1}^{p+1} x_{\ell} = 1\}$
  be the standard $p$-simplex.
  We consider here the case $\Xi_1 = S^p$ and $\Xi_2 = S^p$ --
   the two spaces are, in fact, commensurate in this case, for illustration.
   Let $\bm{\gamma}_i \iid Dirichlet(\1)$ represent $n$ ``objects'' or ``topics''.
   Let $X_{ik} \iid Dirichlet(r\bm{\gamma}_i+\1)$ represent document $i$ in language $k$.
  (Since the $X_{ik}$ take their value in $S^p$, we can think of them as modelling
  (normalized) word count histograms with $p+1$ distinct words.
  $\Xi_1 = \Xi_2 = S^p$ suggests a simplified 1-1 word correspondence model.
  A permutation $\sigma$
  indicating that the 1-1 word correspondence is unknown
  may be applied to the dimensions of one space
  with no alteration to our illustration.)
In this case, $r$ controls what it means to be matched
-- e.g., document translation quality analogy.
If $r$ is large (highly accurate translations), then matched documents $X_{i1}$ and $X_{i2}$
will be probabilistically more similar than $X_{i1}$ and $X_{i'2}$ for $i \neq i'$;
if $r$ is small (rough translations), then ``matched'' doesn't mean much.
Indeed, the limiting case of $r \to \infty$ (point masses) yields ``matched'' means ``identical''
while $r=0$ (recall that $Dirichlet(\1)$ is uniform on the simplex) yields ``matched'' means ``no relationship''.
  Figure \ref{fig:dirichlet}, with $p=2$, provides an illustration wherein matched means quite a lot.
A real data version of this setting with multiple documents per topic
is depicted in Figure \ref{fig:MingSun1},
where three Linguistic Data Consortium (LDC) Enron email message topic classes
are projected into the simplex $S^2$ via
Fisher's Linear Discriminant composed with Latent Semantic Analysis (FLD$\circ$LSA)
(see, e.g., \cite{Berry_2003,Berry_2007,TextMining}).

\subsection{Related Work}

The 2006 David Hand polemic \cite{citeulike:3480824}
argued persuasively that a fundamental issue in statistical inference research and development
-- perhaps {\em the} fundamental issue --
is robustness in the face of test data drawn from a distribution {\em not} the same
as the distribution from which the training data are drawn.
The disparate information fusion described above --
combining multiple spaces with different characteristics --
provides a setting for investigation of related issues.
The recent survey \cite{TransferSurvey} considers a wide range of examples and methodologies addressing this phenomenon
in terms of {\it transfer learning}, {\it domain adaptation}, {\it multitask learning}, etc.
The recent special issue \cite{5714387} 
is devoted entirely to dimensionality reduction via subspace and submanifold learning.
The majority of this article considers the Neyman-Pearson hypothesis testing setting,
which provides clarity through the most straightforward of inference tasks.
In Section \ref{section:classification} we briefly consider a {\em ranking} task.

Our dissimilarity-centric approach is motivated by
the 2005 Pekalska and Duin book \cite{1197035} on the dissimilarity representation for pattern recognition
and the far-reaching success of multidimensional scaling methodologies \cite{T52,T58,CC01,BG05}

Combining information from disparate data sources when the information in the various spaces
is fundamentally incommensurate --
that is, a separate collection of useful features can be extracted from each space
but their interpoint geometry precludes profitable alignment in a common space --
is considered via Cartesian product space embedding in \cite{Maetal-JoC}.

Preliminary development of our joint optimization methodology presented herein,
as well as an application to {\em classification} tasks,
is presented in \cite{MMP-QMDNS}.

\subsection{Summary}

In Section 2 we frame the problem as an optimization problem,
and lay the groundwork for the methodologies proposed in Section 3.
Section 4 illustrates the methodologies with instructive
simulations that illustrate characteristic behavior;
in particular, a simulation involving Dirichlet random variables
sets the stage for the experimental examples on text documents presented in Section 5.
Finally, Section 6 provides discussion and suggestions for several areas of continuing research.

\section{Fidelity and Commensurability}

As suggested in Figure \ref{fig:mm}, our goal is to identify
maps $\rho_k$ taking $\Xi_k$ to $\mathbb{R}^m$ (for some pre-specified $m$)
such that (for $K=2$) the power of the test,
$P[d(\widetilde{\bm{y}}_1,\widetilde{\bm{y}}_2)>c_{\alpha} | H_A: \bm{y}_{1} \nsim \bm{y}_{2}]$,
is large, where the critical value $c_{\alpha}$ is determined by the null distribution of the test statistic
and the allowable Type I error level $\alpha$.

We proceed using $\ell_2$ error for convenience and simplicity;
clearly there is ample reason to consider other error criteria for particular applications.
Similarly, we will assume symmetric dissimilarities $\delta_k$.

The available matched points $\{\bm{x}_{ik}\}$ are used to identify appropriate maps $\rho_k$.
Fidelity is how well the mapping $\bm{x}_{ik} \mapsto \widetilde{\bm{x}}_{ik}$ preserves original dissimilarities.
The within-condition squared {\em fidelity error} is given by
    $$\epsilon^2_{f_{k}} = \frac{1}{{{n}\choose{2}}} \sum_{1 \leq i < j \leq n} (d(\widetilde{\bm{x}}_{ik},\widetilde{\bm{x}}_{jk})-\delta_k(\bm{x}_{ik},\bm{x}_{jk}))^2$$
for each $k$.
If the fidelity error is large, then it is likely that the mapping does not capture aspects of
original data that may be needed for inference.

On the other hand, even if all fidelity errors are small,
inference may fail if
$d(\widetilde{\bm{y}}_1,\widetilde{\bm{y}}_2)$ is large
under the ``matched'' null hypothesis
$H_0: \bm{y}_{1} \sim \bm{y}_{2}$.
Commensurability is how well the mappings preserve matchedness;
the between-condition squared {\em commensurability error} is given by
    $$\epsilon^2_{c_{k_1k_2}} = \frac{1}{n} \sum_{1 \leq i \leq n} (d(\widetilde{\bm{x}}_{ik_1},\widetilde{\bm{x}}_{ik_2})- \delta_{k_1k_2}(\bm{x}_{ik_1},\bm{x}_{ik_2}))^2.$$
Alas, $\delta_{k_1k_2}$ does not exist -- we have no dissimilarity on $\Xi_{k_1} \times \Xi_{k_2}$.
However, the concept of ``matchedness'' suggests that it might be reasonable to set $\delta_{k_1k_2}(\bm{x}_{ik_1},\bm{x}_{ik_2}) = 0$ for all $i,k_1,k_2$,
in which case the commensurability error is the mean squared distance between
matched points -- the same criterion optimized by the Procrustes matching employed below.

There is also between-condition squared {\em separability error} given by
    $$\epsilon^2_{s_{k_1k_2}} = \frac{1}{{{n}\choose{2}}} \sum_{1 \leq i < j \leq n} (d(\widetilde{\bm{x}}_{ik_1},\widetilde{\bm{x}}_{jk_2})- \delta_{k_1k_2}(\bm{x}_{ik_1},\bm{x}_{jk_2}))^2.$$
However, it is less clear how to identify a reasonable stand-in for the $\delta_{k_1k_2}$ terms in this expression.
We will return to this issue when presenting our joint optimization inference methodology proposal in Section \ref{section:omnibus} below.

If all these errors are small -- and if the target dimensionality is low enough
so that estimation variance does not dominate (see e.g.\ \cite{jdm00} Section 3 and \cite{DGL} Figure 12.1) --
then successful inference in the target space may be achievable.
The idea of the joint optimization method proposed in this manuscript (Section \ref{section:omnibus})
is to attempt to minimize all three of these errors simultaneously.

\section{Inference Methodologies}

In this section we present three methodologies for performing our manifold matching inference --
one which focuses on fidelity and is based on multidimensional scaling
and Procrustes matching, one which focuses on commensurability and is based on canonical correlation analysis,
and then our proposal for joint optimization of fidelity and commensurability.

Before proceeding, we briefly review multidimensional scaling, Procrustes matching, and canonical correlation analysis.

Multidimensional scaling (MDS) takes an $n \times n$ dissimilarity matrix $\Delta=[\delta_{ij}]$
and produces a configuration of $n$ points $\widetilde{\bm{x}}_1,\ldots,\widetilde{\bm{x}}_n$ in a target metric space endowed with distance function $d$
such that the collection $\{d(\widetilde{\bm{x}}_i,\widetilde{\bm{x}}_j)\}$ agrees as closely as possible with the original $\{\delta_{ij}\}$
under some specified error criterion; see for instance \cite{T52,T58,CC01,BG05}.
For example, $\ell_2$ (also known as ``raw stress'') MDS minimizes
$\sum_{1 \leq i < j \leq n} (d(\widetilde{\bm{x}}_{i},\widetilde{\bm{x}}_{j})-\delta_{ij})^2$.

Out-of-sample embedding is used throughout this paper -- given a configuration $\{\widetilde{\bm{x}}_i\}_{i=1}^n$
of the training observations
and dissimilarities between test observations and the training observations, the test points are embedded
into the existing configuration so as to be as $\ell_2$-consistent as possible with these dissimilarities.
This out-of-sample embedding can be one at a time, or jointly if the dissimilarities among multiple test observations are also available.
Trosset and Priebe \cite{TP-CSDA-2008-oos} present the out-of-sample methodology appropriate for classical MDS embeddings.
We use raw stress embeddings herein, and the appropriate corresponding out-of-sample methodology is presented in \cite{MPoos}.

Procrustes matching \cite{procrustes1,procrustes2,procrustes3,procrustes4}
takes two matched collections $\widetilde{X}_1$ and $\widetilde{X}^{\prime}_2$ of $n$ points in $\mathbb{R}^m$
and finds the rigid motion transformation which optimally aligns the two collections.
For example, $\ell_2$ Procrustes minimizes the Frobenius norm $\|\widetilde{X}_1 - \widetilde{X}^{\prime}_2Q\|_F$ over all $m \times m$ matrices $Q$ such that $Q^TQ = I$.
(We assume the dissimilarities have been scaled so that a
scaling is not required in the Procrustes mapping.
Thus Q defines a rigid motion mapping $\widetilde{X}^{\prime}_2$ ``onto'' $\widetilde{X}_1$.
We address this issue briefly in Section 6.)

Canonical correlation analysis (CCA) takes a collection $X_1$ of $n_1$ points in $\mathbb{R}^{m_1}$
and a collection $X_2$ of $n_2$ points in $\mathbb{R}^{m_2}$
and finds the pair of linear maps $U_1:\mathbb{R}^{m_1} \to \mathbb{R}$ and $U_2:\mathbb{R}^{m_2} \to \mathbb{R}$
which maximizes the correlation between $\widetilde{X}_1=U_1(X_1)$ and $\widetilde{X}_2=U_2(X_2)$.
Performing $m$ iterations of this procedure in the successive orthogonal subspaces yields a
CCA procedure which maps to $\mathbb{R}^m$. See, for instance, \cite{Hotelling1936,Mardia1980,Hardoon2004}.

Let us now consider these tools as building blocks for manifold matching inference.

\subsection{Procrustes $\circ$ MDS}

Multidimensional scaling yields low-dimensional embeddings.
That is, $\Delta_1 \mapsto \widetilde{X}_1$ and $\Delta_2 \mapsto \widetilde{X}^{\prime}_2$
yields $n \times m$ configurations. 
Procrustes$(\widetilde{X}_1, \widetilde{X}^{\prime}_2)$ yields
$$Q^* = \argmin_{Q^TQ = I} \|\widetilde{X}_1 - \widetilde{X}^{\prime}_2Q\|_F.$$
Given $\delta_k(\bm{y}_{k}, \bm{x}_{ik}) , \ i = 1, \ldots, n$ for $k=1,2$,
out-of-sample embedding of the test data gives $\bm{y}_1 \mapsto \widetilde{\bm{y}}_1, \ \bm{y}_2 \mapsto \widetilde{\bm{y}}'_2$
where the embedded points are chosen so that their distances to $\widetilde{\bm{x}}_{ik}$ agree as closely as possible with the available dissimilarities.
Using the rigid motion transformation obtained in the Procrustes step,
both $\widetilde{\bm{y}}_1$ and $\widetilde{\bm{y}}_2 = ((\widetilde{\bm{y}}'_2)^T Q^*)^T$ are in $\mathbb{R}^m$ with same
coordinate system.
Thus inference may proceed by rejecting for large values of $d(\widetilde{\bm{y}}_1,\widetilde{\bm{y}}_2)$.
We dub this separate embedding approach 
``Procrustes composed with multidimensional scaling'', or ``{\it p}$\circ${\it m}''.

From an inspection of the raw stress multidimensional scaling criterion function,
it follows immediately that the $\Delta_k \mapsto \widetilde{X}_k$ mappings minimize fidelity error.
Thus we have established the following result:
\\

  Theorem 1: {\it p}$\circ${\it m} optimizes fidelity {\em without regard for commensurability}.
\\

That is, the maps $\rho_k$ are identified separately, with no concern for whether
the commensurability optimization in
the Procrustes step will be able to provide a good alignment.

\subsection{Canonical Correlation}

Since canonical correlation begins with Euclidean data,
the first step of this methodology necessarily involves multidimensional scaling.
This appears similar to Procrustes $\circ$ MDS above,
but in this case no attempt is made to achieve meaningful dimensionality reduction.
Multidimensional scaling yields high-dimensional embeddings, $\Delta_1 \mapsto X_1'$ and $\Delta_2 \mapsto X_2'$,
but in this case these maps are to the highest-dimensional space possible, $\mathbb{R}^{n-1}$ in general.
Canonical correlation finds linear maps to $\mathbb{R}^m$, $U_1: X_1' \mapsto \widetilde{X}_1$ and $U_2: X_2' \mapsto \widetilde{X}_2$, to maximize correlation.
Again, out-of-sample embedding yields $(n-1)$-dimensional points
$\bm{y}_1 \mapsto {\bm{y}}'_1, \ \bm{y}_2 \mapsto {\bm{y}}'_2$.
Then $\widetilde{\bm{y}}_1 = U_1^T {\bm{y}}'_1$ and $\widetilde{\bm{y}}_2 = U_2^T {\bm{y}}'_2$ can be directly compared.
An investigation of the correlation criterion function shows that
the CCA maps $U_1$ and $U_2$ minimize commensurability error, subject to linearity.
Thus there is no need for Procrustes in this case,
and once again inference may proceed:
reject for large values of $d(\widetilde{\bm{y}}_1,\widetilde{\bm{y}}_2)$.
We dub this approach ``{\it cca}''.

From the equivalence of the correlation objective function and commensurability error, we have established the following result:
\\

  Theorem 2: {\it cca} optimizes commensurability {\em without regard for fidelity}.
\\

That is, the maps $\rho_k$ are identified jointly, but with no concern for fidelity of the individual embeddings
(beyond linearity).

\subsection{Omnibus Embedding}\label{section:omnibus}

In response to the optimization objectives of the two methodologies presented above --
one considering fidelity only and the other considering commensurability only --
we develop an omnibus embedding methodology explicitly focused on the joint optimization of fidelity and commensurability.

\begin{figure}[h]
  \begin{center}
    \includegraphics[scale=1.6]{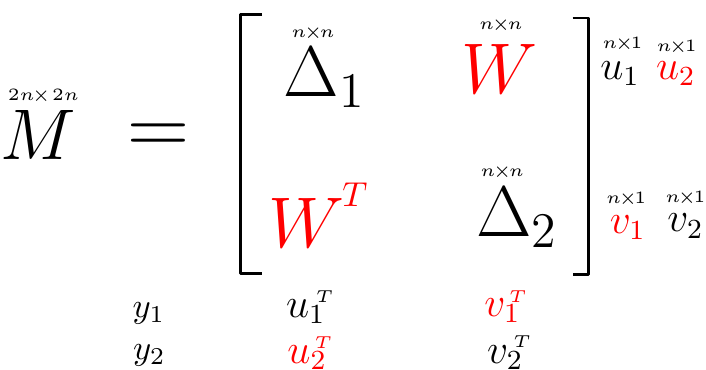}
    \caption{Depiction of the $2n \times 2n$ omnibus dissimilarity matrix $M$, including imputed dissimilarities
    $W=[\delta_{12}(\bm{x}_{i1},\bm{x}_{j2})]$ and out-of-sample test data $\bm{y}_{1}, \bm{y}_{2}$.}\label{fig:M}
  \end{center}
  \end{figure}

Under the ``matched'' assumption, we {\em impute} dissimilarities $W = [\delta_{12}(\bm{x}_{i1},\bm{x}_{j2})]$
to obtain a $2n \times 2n$ {\em omnibus dissimilarity matrix} $M$. See Figure \ref{fig:M},
which depicts $M$ as a block matrix consisting of the $n \times n$ dissimilarities matrices $\Delta_1$ and $\Delta_2$ on
the diagonal and $W$ as the $n \times n$ off-diagonal block.
(This generalizes immediately to $K>2$.)
As discussed above,
it seems reasonable under $H_0$ to set the diagonal elements $\delta_{k_1k_2}(\bm{x}_{ik_1},\bm{x}_{ik_2})$ of $W$ to zero.
(Notice, however, that
$\delta_{k_1k_2}(\bm{x}_{ik_1},\bm{x}_{ik_2})=0$ for $k_1 \neq k_2$
is not necessarily ``truth.'' For instance, the Dirichlet setting of Section \ref{section:dirichlet} with $r < \infty$
would have non-zero elements for $diag(W)$. Still, this ``shrinkage'' of $diag(W)$ to zero seems reasonable.)
As for the off-diagonal elements of $W$, we argue that either leaving them as missing data
unused in the subsequent optimization or letting $W=(\Delta_1 + \Delta_2)/2$ are reasonable suggestions;
we will return to this imputation issue later.
Once we have settled on $W$, our approach considers MDS embedding of $M$ as $2n$ points in $\mathbb{R}^{m}$ --
zeros on the diagonal of $W$ act to force matched points to be embedded near each other.
It is clear that raw stress MDS applied to $M$ has as its objective function precisely
$\epsilon^2_{f_{1}}$ + $\epsilon^2_{f_{2}}$ + $\epsilon^2_{c_{12}}$ + $\epsilon^2_{s_{12}}$.
If $diag(W)=0$ and the off-diagonal elements are treated as missing and ignored in the optimization,
then this objective function reduces to a consideration of just fidelity and commensurability.

  Let $u_{i1} = \delta_1(\bm{y}_{1}, \bm{x}_{i1})$ and $v_{i2} = \delta_2(\bm{y}_{2}, \bm{x}_{i2})$.
  Under $H_0$, impute $v_{i1} = \delta_{12}(\bm{y}_1,\bm{x}_{i2})$ and $u_{i2} = \delta_{12}(\bm{y}_2,\bm{x}_{i1})$
  via $\bm{v}_1 = \bm{u}_2 = (\bm{u}_1 + \bm{v}_2)/2$.
  Out-of-sample embedding of $(\bm{u}_1^T, \bm{v}_1^T)^T$ and $(\bm{u}_2^T, \bm{v}_2^T)^T$ yields $\widetilde{\bm{y}}_1$ and $\widetilde{\bm{y}}_2$.
  Reject for large values of $d(\widetilde{\bm{y}}_1,\widetilde{\bm{y}}_2)$.
We dub this omnibus embedding approach for joint optimization of fidelity and commensurability ``{\it jofc}''.

Obviously, the choice of $W$ is key for this joint optimization.
Also, note that weights can be incorporated into the MDS optimization criterion;
this weighting can become quite elaborate,
but in its simplest form it yields a more general tradeoff between fidelity and commensurability via $\omega (\epsilon^2_{f_{1}} + \epsilon^2_{f_{2}})$ + $(1-\omega) \epsilon^2_{c_{12}}$.

\section{Illustrative Simulation}

In this section we present an illustrative Dirichlet simulation
which helps to elucidate when and why our joint optimization methodology is superior to
either version of separate optimization.

\subsection{Dirichlet Product Model}

We describe a probability model with parameters $p,q,r,a$, and $K$.

Let $\Xi_k = S^{p+q}$, $k=1,2$.
Here the simplex $S^p$ encodes ``signal'' and
the simplex $S^q$ encodes ``noise''.
That is, on $S^p$ we let $\bm{\gamma}_i \iid Dirichlet(\1)$
and mutually independent $X^1_{ik} \sim Dirichlet(r\bm{\gamma}_i+\1)$
(signal, as in Section \ref{section:dirichlet})
while on $S^q$ we let $X^2_{ik} \iid Dirichlet(\1)$ (pure noise).
For $a \in [0,1]$, let $X_{ik} = [(1-a) X^1_{ik} , a X^2_{ik}]$ --
the concatenation of (weighted) signal and noise dimensions.
The resultant distribution for $(X_{i1},\cdots,X_{iK})$ is denoted by $F_{p,q,r,a,K}$,
and $F_{p,q,r,a,K|\bm{\gamma}_1,\cdots,\bm{\gamma}_n}$ denotes the distribution conditional on the location of the $\bm{\gamma}_i$.

\subsection{Testing}

For each of $n_{mc}$ Monte Carlo replicates ($n_{mc}=1000$ in the simulations), we generate $n$ matched pairs
according to the Dirichlet product model distribution $F_{p,q,r,a,K=2}$
by first generating $\bm{\gamma}_1,\ldots,\bm{\gamma}_n$
and then, conditional on the collection $\{\bm{\gamma}_i\}$, generating the matched pair $(X_{i1},X_{i2})$.
Embeddings are defined for each of the three competing methodologies
based on this matched training data.
For each test datum under $H_0$, one new $\bm{\gamma}$ is generated,
a matched pair is generated, out-of-sample embedding is performed,
and the statistic $T=d(\widetilde{\bm{y}}_1,\widetilde{\bm{y}}_2)$ is calculated;
this is repeated $s$ times independently ($s=1000$ in the simulations)
and the critical value $c_{\alpha}$ for the allowable Type I error level $\alpha$
is determined based on the Monte Carlo estimate of null distribution of $T$.
Then {\em un}matched pairs are generated, out-of-sample embedding is performed,
and the statistic $T$ is calculated for test data under $H_A$;
this provides an estimate of the conditional power
$P[d(\widetilde{\bm{y}}_1,\widetilde{\bm{y}}_2)>c_{\alpha} | H_A,\bm{\gamma}_1,\ldots,\bm{\gamma}_n]$.

We perform $n_{mc}$ Monte Carlo replicates to integrate out the $\bm{\gamma}_1,\ldots,\bm{\gamma}_n$,
yielding comparative power estimates.
We also investigate conditional power for particular collections $\{\bm{\gamma}_i\}$,
in order to better understand
precisely when and why our joint optimization methodology is superior to
either version of separate optimization.

\subsection{Results}

\begin{figure}[h]
\begin{center}
  \includegraphics[height=12cm, width=14.4cm,angle=0]{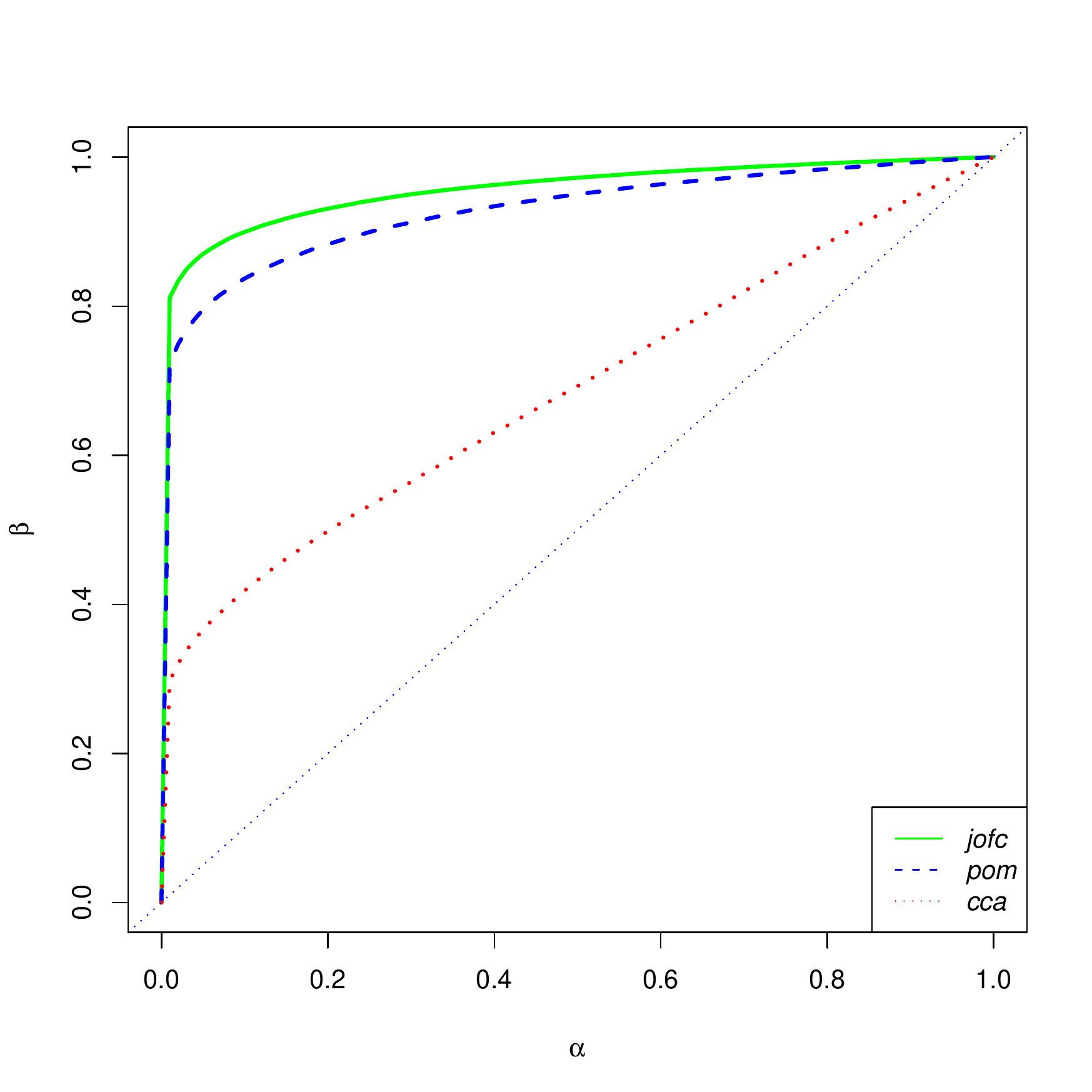}
\caption{
Dirichlet product model
simulation results
plotting the Type I error level $\alpha$ against power $\beta = P[d(\widetilde{\bm{y}}_1,\widetilde{\bm{y}}_2)>c_{\alpha} | H_A]$,
indicating that {\em jofc} is superior to both {\it p}$\circ${\it m} and {\em cca}. See text for description.
}\label{fig:simrocD}
\end{center}
\end{figure}

Figure \ref{fig:simrocD} presents results
from our Dirichlet product model. $K=2$, with $p=3,q=3,r=100,a=0.1$.
The target dimension is $m=2$. We use $n=100$.
The allowable Type I error level $\alpha$ is plotted against power $\beta = P[d(\widetilde{\bm{y}}_1,\widetilde{\bm{y}}_2)>c_{\alpha} | H_A]$.
The results are based on $n_{mc}=1000$ Monte Carlo replicates with $s=1000$;
the differences in the curves are statistically significant.
In this case, {\em jofc} with $W=(\Delta_1 + \Delta_2)/2$ is superior to both {\it p}$\circ${\it m} and {\em cca}.

\subsection{Analysis}

The Dirichlet product model
is designed specifically to illustrate when and why {\em jofc} is superior to both {\it p}$\circ${\it m} and {\em cca}
in terms of fidelity and commensurability.

If $q$ is large with respect to the target dimensionality $m$,
then with high probability {\em cca} will identify a $m-$dimensional subspace in the ``noise'' simplex $S^q$ with spurious correlation.
This phenomenon requires only that $a>0$.
In this event, the out-of-sample embedding will produce arbitrary $\widetilde{\bm{y}}_1$ and $\widetilde{\bm{y}}_2$,
even under $H_0$.
Thus the null distribution of the test statistic will be inflated by these spurious correlations.
If the allowable Type I error level is smaller than the probability of inflation,
then the power of the {\em cca} method will be negatively affected.

\begin{figure}
  \begin{center}
    \includegraphics[height=13.2cm, width=15.84cm,angle=0]{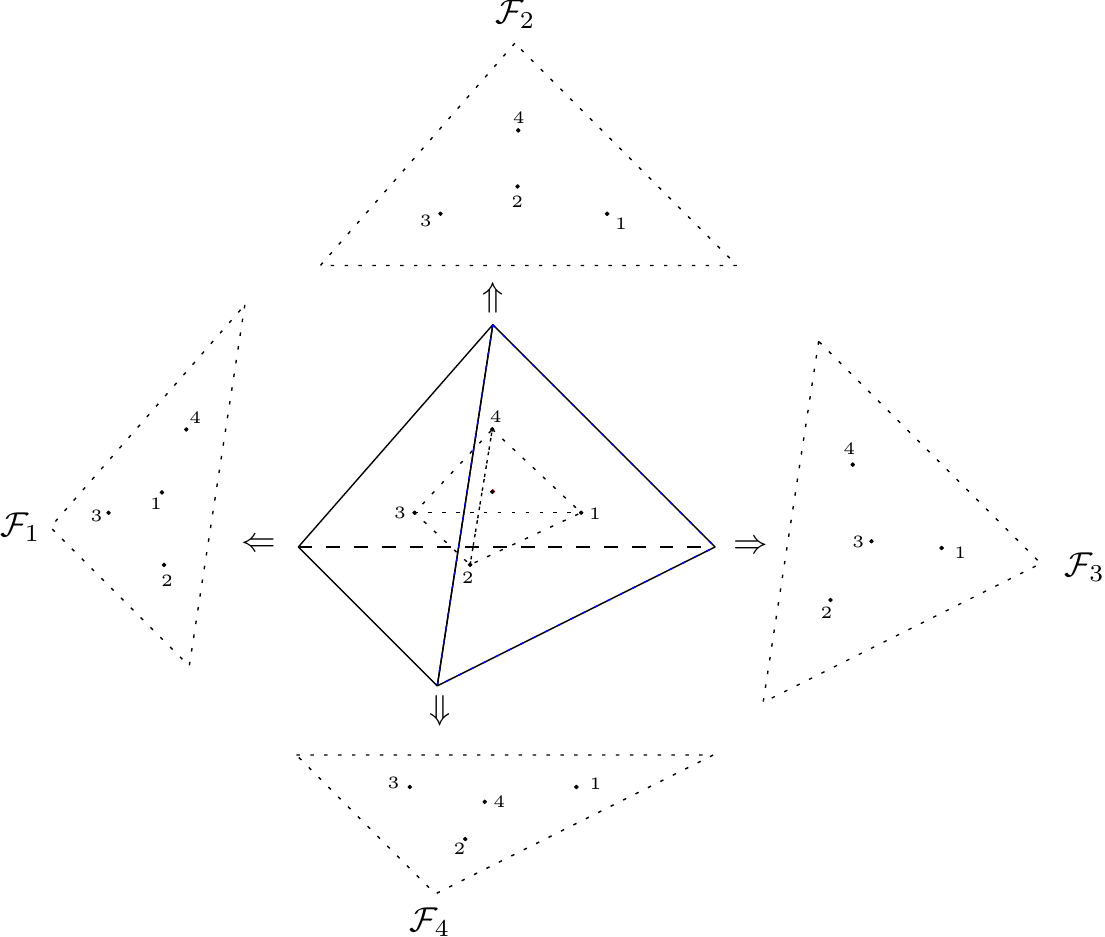}
    \caption{Idealization of the incommensurability phenomenon:
    for a symmetric collection $\{\bm{\gamma}_1,\bm{\gamma}_2,\bm{\gamma}_3,\bm{\gamma}_4\}$ in the simplex $S^3$,
    all four of the facet projections have the same fidelity and are geometrically incommensurable with one another.}\label{fig:incomm}
  \end{center}
  \end{figure}

If $a$ is small and $m \leq p$, then with high probability the $m-$dimensional subspaces identified by the MDS step
will come from the ``signal'' simplex $S^p$.
If $m<p$, then with positive probability, these two subspaces,
identified separately in {\it p}$\circ${\it m},
will be geometrically incommensurate (see Figure \ref{fig:incomm}).
Thus the null distribution of the test statistic will be inflated by these incommensurate cases.
If the allowable Type I error level $\alpha$ is smaller than the probability of inflation,
then the power of the {\it p}$\circ${\it m} method will be negatively affected.

For large $q$ and small $a$, the two phenomena described above occur in the same model.
The {\it jofc} method is not susceptible to either phenomenon:
incorporating fidelity into the objective function obviates the spurious correlation phenomenon, and
incorporating commensurability into the objective function obviates the geometric incommensurability phenomenon.
Thus we can establish that,
for a range of Dirichlet product model distributions, {\em jofc} is superior to both {\it p}$\circ${\it m} and {\em cca}.
\\

Theorem 3: Let $m \in \{1,\cdots,\min\{p-1,q\}\}$, $a \in (0,1/2)$, and $r \in (0,\infty)$.
Then for large $q$, small $a$, and large $r$,
there exists allowable Type I error level $\alpha > 0$
such that the Dirichlet product model distribution $F_{p,q,r,a,K=2}$ with target dimensionality $m$
yields power $\beta_{jofc} > \max\{\beta_{\protect{{\it p}\circ{\it m}}},\beta_{cca}\}$,
where power $\beta = P[d(\widetilde{\bm{y}}_1,\widetilde{\bm{y}}_2)>c_{\alpha} | H_A]$
for the various testing methodologies {\em jofc}, {\it p}$\circ${\it m}, and {\em cca}.
\\

Proof:
Let $b_1$ denote the probability that {\em cca}
suffers from the spurious correlation phenomenon,
and let $b_2$ denote the probability that {\it p}$\circ${\it m}
suffers from the geometric incommensurability phenomenon.
Then $q \gg p$ implies that {\em cca} suffers from the spurious correlation phenomenon
with high probability and thus $b_1 \approx 1$ and $\beta_{cca} \approx \alpha$.
For $a \approx 0$ and $r$ sufficiently large,
{\em jofc} and {\it p}$\circ${\it m} identify approximately the same embeddings
{\em except} for the cases in which
 {\it p}$\circ${\it m} suffers from the incommensurability phenomenon.
 Thus the null distribution of $T=d(\widetilde{\bm{y}}_1,\widetilde{\bm{y}}_2)$ for {\em jofc} is approximately point mass at zero
while the null distribution of $T$ for {\it p}$\circ${\it m} has $b_2$ mass $\gg 0$.
Hence $\alpha \approx b_2/2$ yields
$\beta_{jofc} \approx 1$
while $\beta_{\protect{{\it p}\circ{\it m}}} \approx 1/2$.$_\blacksquare$
\\

Delving into our simulation results via investigation of conditional power
$P[d(\widetilde{\bm{y}}_1,\widetilde{\bm{y}}_2)>c_{\alpha} | H_A,\bm{\gamma}_1,\ldots,\bm{\gamma}_n]$,
it is apparent that the superiority of {\em jofc}
is indeed due to occurrences of the phenomena described above --
individual Monte Carlo replicates (particular selections of the $\{\bm{\gamma}_i\}$, essentially)
are identified in which the spurious correlation phenomenon causes poor performance for {\it cca}
or the incommensurability phenomenon causes poor performance for {\it p}$\circ${\it m}
but in which {\em jofc} is unaffected.

We note that the Dirichlet product model introduced here as an aid in understanding
when and why {\em jofc} is superior to both {\it p}$\circ${\it m} and {\em cca}
does in fact (loosely) model general high-dimensional real data scenarios:
many dimensions consisting mostly of noise along with a few signal dimensions.

\subsection{Gaussian Model}

A Gaussian model, analogous to the Dirichlet product model investigated above,
is constructed here to provide a sense of the generality of models
with many dimensions consisting mostly of noise along with a few signal dimensions.

We consider $p$-dimensional means $\bm{\mu}_{i}\iid\mathcal{N}\left(\vec{0},I_p\right)$, $i=1,\cdots,n$,
analogous to the $\bm{\gamma}_i$ from the Dirichlet model.
Matchedness arises from independent $X^1_{ik} \sim\mathcal{N}\left(\bm{\mu}_{i},r^{-1}I_p\right)$, $i=1,\ldots,n$, $k=1,\ldots K$,
for $r \in (0,\infty)$;
as $r$ increases, the degree of matchedness increases.
As before, we have $q$-dimensional ``noise'' vectors
$X^2_{ik} \iid \mathcal{N}\left(\vec{0},I_{q}\right)$.
Again, for $a \in [0,1]$, $X_{ik} = [(1-a) X^1_{ik} , a X^2_{ik}]$
represents the concatenation of (weighted) signal and noise dimensions.
As with the Dirichlet product model,
both the spurious correlation phenomenon
and the geometric incommensurability phenomenon are present in this Gaussian model.

\begin{figure}[h]
\begin{center}
\includegraphics[height=12cm, width=14.4cm,angle=0]{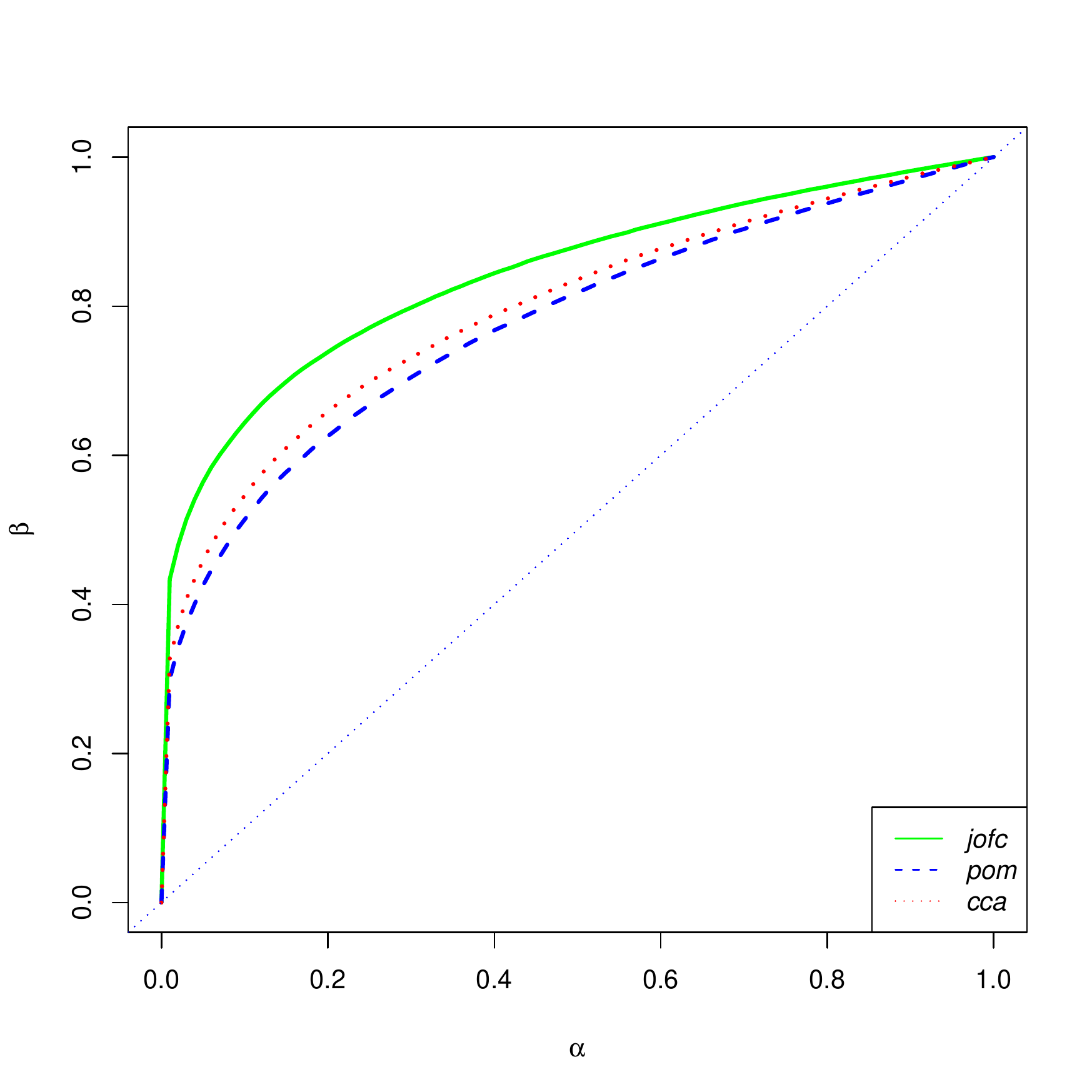}
\caption{
Gaussian model
simulation results
plotting the Type I error level $\alpha$ against power $\beta = P[d(\widetilde{\bm{y}}_1,\widetilde{\bm{y}}_2)>c_{\alpha} | H_A]$,
indicating {\em jofc} is superior to both {\it p}$\circ${\it m} and {\em cca},
entirely analogous to those presented for the Dirichlet product model in Figure \ref{fig:simrocD}.
}\label{fig:simrocG}
\end{center}
\end{figure}

Figure \ref{fig:simrocG} presents simulation results for this Gaussian model,
entirely analogous to those depicted in Figure \ref{fig:simrocD}.

\section{Experimental Results}

\subsection{Testing}\label{section:testing}

A collection of documents $\{\bm{x}_{i1}\}_{i=1}^n$ are collected from the English Wikipedia,
corresponding to the directed 2-neighborhood of the document ``Algebraic Geometry.''
This yields $n=1382$ and, through Wikipedia's own 1-1 correspondence,
the associated French documents $\{\bm{x}_{i2}\}_{i=1}^n$.
For dissimilarity matrices $\Delta_k$, $k=1,2$, we use the Lin \& Pantel discounted mutual information \cite{LinPantel1,LinPantel2}
and cosine dissimilarity $\delta_k(\bm{x}_{ik},\bm{x}_{jk}) = 1-(\bm{x}_{ik} \cdot \bm{x}_{jk})/(\|\bm{x}_{ik}\|_2 \|\bm{x}_{jk}\|_2)$.

Our results are obtained by repeatedly randomly holding out four documents -- two matched pairs --
and identifying the embeddings via {\em cca}, {\it p}$\circ${\it m}, and {\em jofc}
based on the remaining $n=1380$ matched pairs.
The two sets of held-out matched pairs are used as $\bm{y}_{1}$ and $\bm{y}_{2}$,
via out-of-sample embedding, to estimate the null distribution of the
test statistic $T=d(\widetilde{\bm{y}}_1,\widetilde{\bm{y}}_2)$.
This allows us to estimate critical values for any specified Type I error level.
Then the two sets of held-out {\em un}matched pairs are used as $\bm{y}_{1}$ and $\bm{y}_{2}$,
via out-of-sample embedding, to estimate power.
Target dimensionality $m$
is determined by the Zhu and Ghodsi automatic dimensionality selection method \cite{ZhuGhodsi},
resulting in $m=6$ for this data set.

\begin{figure}[h]
\begin{center}
\includegraphics[height=12cm, width=14.4cm,angle=0]{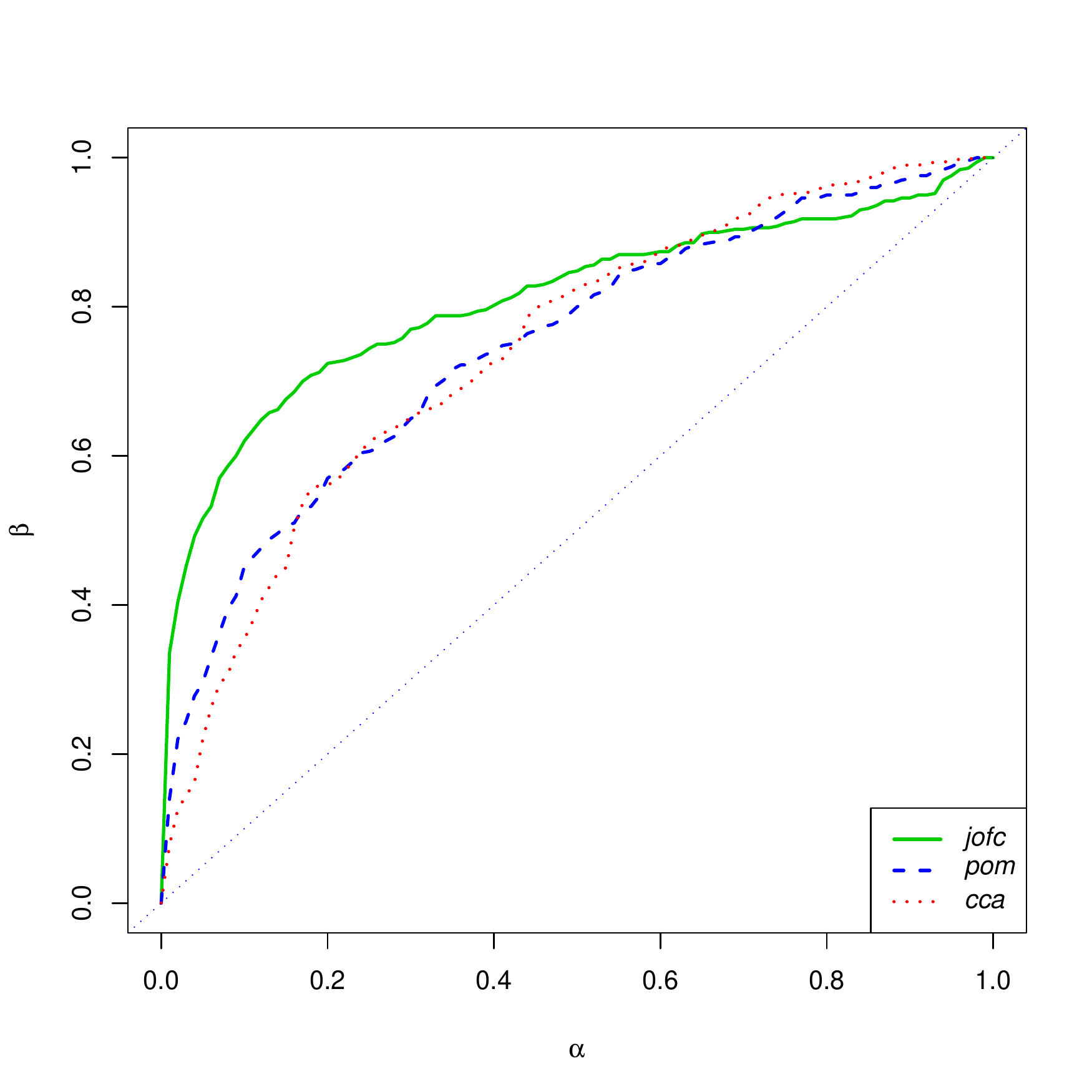}
\caption{
Experimental results on English/French Wikipedia documents
plotting the Type I error level $\alpha$ against power $\beta = P[d(\widetilde{\bm{y}}_1,\widetilde{\bm{y}}_2)>c_{\alpha} | H_A]$,
indicating {\em jofc} is superior to both {\it p}$\circ${\it m} and {\em cca}.
See text for description.
}\label{fig:exproc}
\end{center}
\end{figure}

Figure \ref{fig:exproc} plots the allowable Type I error level against power.
These experimental results indicate that {\em jofc} is superior to both {\it p}$\circ${\it m} and {\em cca},
and are entirely analogous to the simulation results presented above.

\subsection{Ranking}\label{section:classification}

Here we consider a {\em ranking} task
in which matched training data exists in disparate spaces $\Xi_1$ and $\Xi_2$,
but test observation $\bm{y}_{2}$
will be observed in space $\Xi_{2}$.
The task is to find the match for $\bm{y}_{2}$
amongst a candidate collection $\mC = \{\bm{y}_{11}, \cdots, \bm{y}_{z1}\} \subset \Xi_1$
of $z>1$ possibilities.
Using the training set of matched observations,
we identify the embeddings via {\em cca}, {\it p}$\circ${\it m}, and {\em jofc},
and out-of-sample embedding then yields $\widetilde{\bm{y}}_{2}$
and $\widetilde{\mC} = \{\widetilde{\bm{y}}_{11}, \cdots, \widetilde{\bm{y}}_{z1}\}$.
The {\em rank} $r^*$ of the one true match to $\bm{y}_{2}$
amongst the candidate collection $\mC$
in terms of $\{d(\widetilde{\bm{y}}_{\zeta1},\widetilde{\bm{y}}_{2})\}_{\zeta=1}^z$
is our measure of performance;
$r^* = 1$ represents perfect performance,
$r^* = z/2$ represents chance, and
$r^* = z$ is the worst possible.


For this experiment we consider a different collection of Wikipedia documents:
all English/Persian (Farsi) matched pairs
(matched, again, through Wikipedia's own 1-1 correspondence)
for which both documents in the pair contain
at least 500 total words and at least 100 distinct words.
There are 2448 such pairs.
(The word-count restrictions are to ensure that the documents are legitimate articles,
rather than ``stubs'' -- place-holders for future articles on the topic.)

\begin{figure}[h]
\begin{center}
\includegraphics[height=12cm, width=14.4cm,angle=0]{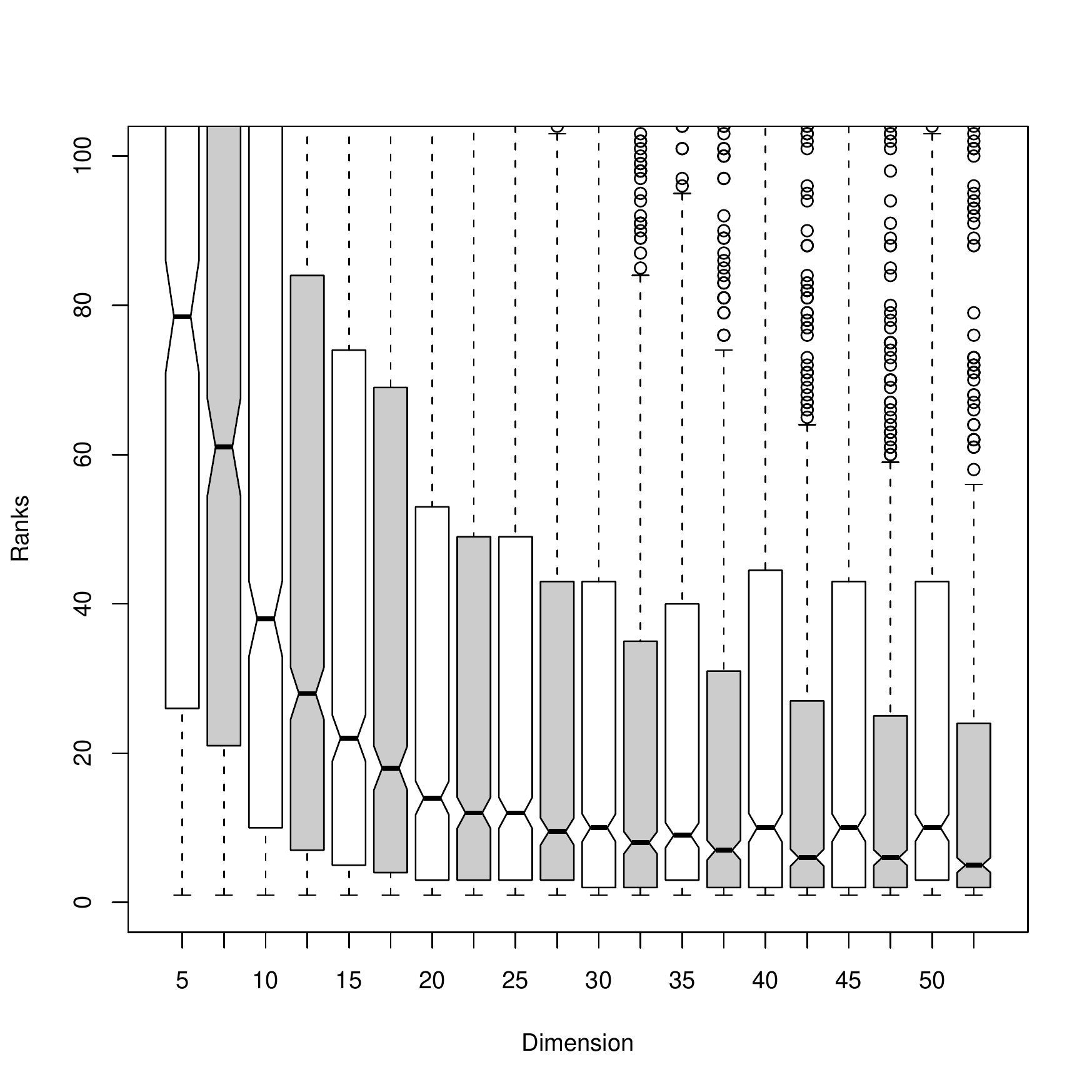}
\caption{
Comparative rank experimental results
depicting the rank $r^*$
of the one true match to test observation $\bm{y}_{2}$
amongst the candidate collection $\mC$
in terms of $\{d(\widetilde{\bm{y}}_{\zeta1},\widetilde{\bm{y}}_{2})\}_{\zeta=1}^z$
as a function of target dimension $m$.
For each $m \in \{5,10,15,\cdots,50\}$, there are two boxplots.
These results indicate that {\em jofc} (gray) is superior to {\it p}$\circ${\it m} (white) on this data set.
With $z=1000$, both methods perform much better than chance ($r^*=z/2$),
although performance does not achieve perfection ($r^*=1$).
See text for description.
}\label{fig:rawz}
\end{center}
\end{figure}

\begin{figure}[h]
\begin{center}
\includegraphics[height=12cm, width=14.4cm,angle=0]{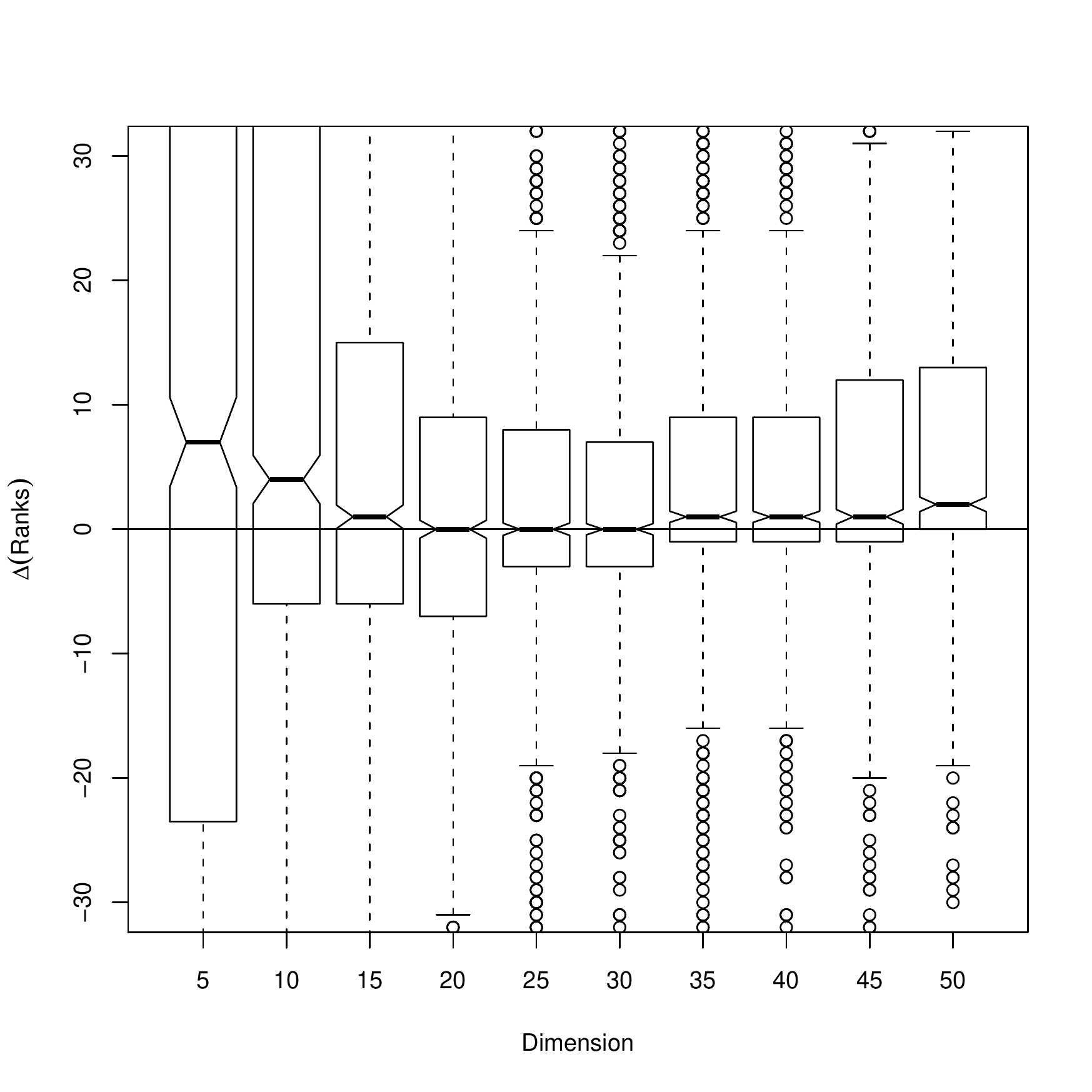}
\caption{
Comparative rank experimental results
depicting difference in ranks
$r^*_{\protect{{\it p}\circ{\it m}}} - r^*_{\protect{\it jofc}}$;
differences greater than 0 indicate {\em jofc} superiority.
See text for description.
}\label{fig:rawdiffz}
\end{center}
\end{figure}

Figures \ref{fig:rawz} and \ref{fig:rawdiffz} present notched boxplot experimental results
wherein we repeatedly hold out $z=1000$ matched pairs from the training set.
(Recall that non-overlapping notches implies a statistically significant difference of means.)
Figure \ref{fig:rawz} depicts $r^*$ as a function of target dimension $m$
for {\em jofc} (gray) and {\it p}$\circ${\it m} (white).
Performance improves for both methods as $m$ increases from 5 to 25, with {\em jofc} superior.
Performance levels off after $m=30$
(and degrades significantly for $m>50$).
Figure \ref{fig:rawdiffz} depicts difference in ranks,
$r^*_{\protect{{\it p}\circ{\it m}}} - r^*_{\protect{\it jofc}}$;
differences greater than 0 indicate {\em jofc} superiority.

\section{Discussion and Conclusions}

We have presented a complete methodological core
for manifold matching via joint optimization of fidelity and commensurability
and comprehensive comparisons with either version of separate optimization.
Continuing research includes comparison with other standard competing methodologies,
variations and generalizations of our omnibus embedding methodology,
and further theoretical developments.

Here we discuss a few of the most pressing issues.

\subsubsection*{$K>2$ Conditions}

It is straightforward to generalize the omnibus dissimilarity matrix $M$ to the case of $K>2$ conditions.

\subsubsection*{Pre-Scaling the $\Delta_k$}

The {\em scale} of the various dissimilarities has been assumed to be consistent.
For Dirichlet data, this assumption is warranted;
however, pre-scaling of the $\Delta_k$ prior to constructing $M$ is imperative for the general case.

\subsubsection*{MDS Objective}

Our omnibus embedding methodology can be employed with MDS criteria other than raw stress;
the $\ell_2$ criterion provides direct correspondence to fidelity and commensurability.
Weighted $\ell_2$ is straightforward.
Other MDS minimization objectives have been studied in depth,
and should in particular circumstances provide superior performance.

\subsubsection*{Imputation of $W$}

It seems reasonable under $H_0$ to set the diagonal elements $\delta_{k_1k_2}(\bm{x}_{ik_1},\bm{x}_{ik_2})$ of $W$ to zero.
Recall, however, that this is not necessarily ``truth;''
the Dirichlet setting of Section \ref{section:dirichlet} with $r < \infty$
would have non-zero elements for $diag(W)$.
Still, this shrinkage of $diag(W)$ to zero seems reasonable.
However, there may be cases for which imputing non-zero values would be appropriate;
for example, if information is available suggesting that some matchings are unreliable,
then it might be advantageous to use larger values for these matchings.

As for the off-diagonal elements of $W$, we have argued that either leaving them as missing data
unused in the subsequent optimization or letting $W=(\Delta_1 + \Delta_2)/2$ are reasonable suggestions.
We believe that more elaborate imputation should provide superior performance.
In particular, it seems clear that choosing $\lambda\in[0,1]$
and setting $W=\lambda\Delta_1+(1-\lambda)\Delta_2$ or
$W=(\lambda\Delta_1^2+(1-\lambda)\Delta_2^2)^{1/2}$
will be preferable in certain circumstances.

\subsubsection*{Model Selection: The Choice of Target Dimensionality $m$}

We have assumed throughout that $\mathcal{X} = \mathbb{R}^m$
for some pre-specified target dimension $m$.
First, we note that, in general, embedding into target spaces other than Euclidean is possible
and sometimes productive.
More pressing is the necessity, in many applications, for data-driven choice of target dimension.
This is in general a vexing model selection task -- the bias-variance trade-off.
Of course, $m=1$ generally induces significant model bias
and $m=n-1$ generally admits excessive estimation variance,
as characterized in \cite{DGL} Figure 12.1.
Many dimensionality selection methods based on the principle of diminishing returns in terms of variance explained
are available -- in Section \ref{section:testing} we made use of the method proposed in \cite{ZhuGhodsi},
and in \ref{section:classification} we presented results as a function of $m$.
A dimensionality selection methodology specifically designed for use with our omnibus embedding methodology is of significant interest.

One illustrative point in this regard is that the general commensurate-space approach considered throughout
this article -- for all three approaches {\em jofc}, {\it p}$\circ${\it m}, and {\em cca} --
adds a further complication with respect to identification of optimal target dimension:
the optimal target dimension $m_k^*$ for the various $\Delta_k$ will not the be same.
This adds to the degree of difficulty in designing methods for identifying
the optimal common-space target dimension $m^*$.

\subsubsection*{Learning the $\pi_k$}

We have assumed that the maps $\pi_k$ from object space $\Xi$
to the conditional spaces $\Xi_k$ are fixed (see Figure \ref{fig:mm}).
Indeed, $\Xi$ and the $\pi_k$ have been treated as notional only.
In some circumstances, it may be possible to use performance analyses
to glean information concerning the induced conditional distributions
and profitably adjust the $\pi_k$, in a manner analogous to fusion frames \cite{FF}.

\subsubsection*{{\em Fast} Omnibus Embedding}

Out-of-sample embedding of test data precludes re-learning the mappings for each inference.
More importantly,
it is straightforward to make a version of our omnibus embedding methodology fast ($O(n)$).
Making an {\em effective} fast version requires numerous methodological choices
for various stages of {\em jofc}.

\subsubsection*{Commensurability Error vs Hausdorff Distance on $G_{p,m}$}\label{section:Gr}


In the simple setting of Euclidean spaces $\Xi_k$,
the {\it p}$\circ${\it m} methodology yields two elements of the Grassmann space $G_{p,m}$ of $m$-dimensional subspaces of $\Real^{p}$.
This space is a manifold under the Hausdorff distance
$2\sin(\theta/2)$, where $\theta$ is the canonical angle between subspaces \cite{1103435}.
Under special conditions the Hausdorff distance between {\it p}$\circ${\it m}'s two subspaces and
the commensurability error between their respective embeddings are closely related.

\begin{figure}[h]
  \begin{center}
    \includegraphics[height=12cm, width=14.4cm,angle=0]{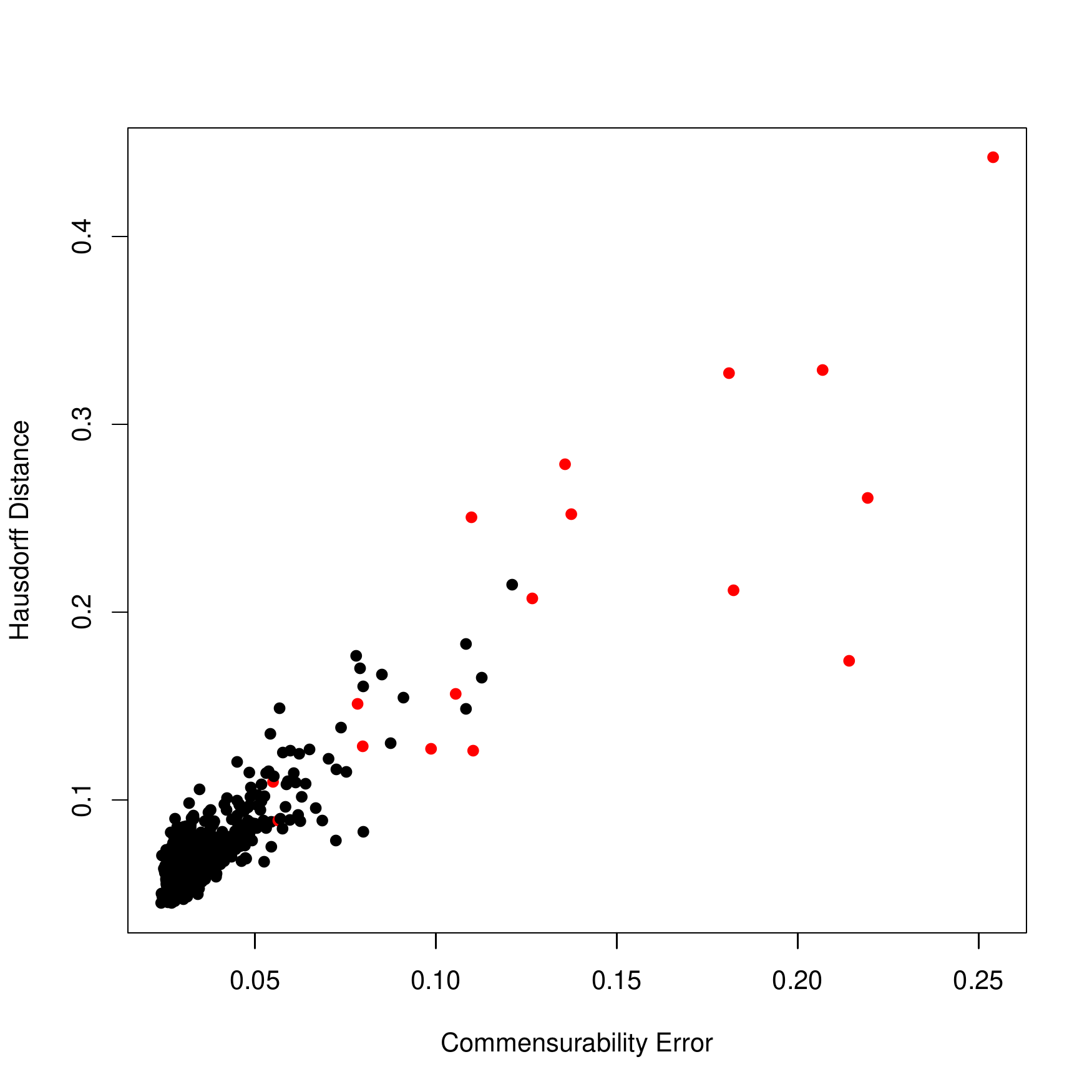}
    \caption{Commensurability error and Hausdorff distance on the Grassmannian Manifold
    for our Dirichlet product model simulation (Figure \ref{fig:simrocD}).
    Strong correlation is evident.
    Furthermore, the red points represent replicates for which the conditional power
	$P[d(\widetilde{\bm{y}}_1,\widetilde{\bm{y}}_2)>c_{\alpha} | H_A,\bm{\gamma}_1,\ldots,\bm{\gamma}_n]$
    is low --
    predominantly those replicates for which Hausdorff distance and commensurability error are large.}\label{fig:grass}
  \end{center}
  \end{figure}

See Figure \ref{fig:grass} for a first example, from the Dirichlet product model simulation presented in
Figure \ref{fig:simrocD}. Each point in Figure \ref{fig:grass} represents a Monte Carlo replicate.
We note that the Hausdorff distance between {\it p}$\circ${\it m}'s two subspaces and
the commensurability error between their respective embeddings are strongly correlated.
Furthermore, the red points represent replicates for which the conditional power
$P[d(\widetilde{\bm{y}}_1,\widetilde{\bm{y}}_2)>c_{\alpha} | H_A,\bm{\gamma}_1,\ldots,\bm{\gamma}_n]$ is low --
predominantly those replicates for which Hausdorff distance and commensurability error are large.
This demonstrates the effect of the incommensurability phenomenon on {\it p}$\circ${\it m}.
The {\em jofc} embeddings are not subject to this deleterious phenomenon.

Additional investigations concerning the superiority of {\em jofc} to {\it p}$\circ${\it m}
due to the incommensurability phenomenon involve this relationship between Hausdorff distance and commensurability error.
Significantly more involved investigations are required when,
as is the case for proper text document analysis, one uses a more appropriate dissimilarity
(Hellinger distance, or more generally $\alpha$-divergence) on the simplex.

\subsubsection*{Three-Way MDS}

Three-way MDS (see, for instance, \cite{BG05})
addresses a problem superficially similar to joint optimization of fidelity and commensurability,
in which a single configuration and two transformation matrices
are identified from two dissimilarity matrices $\Delta_1,\Delta_2$.
It may be of interest to compare and contrast our omnibus embedding methodology
with various instantiations of three-way MDS -- particularly the identity model
presented in \cite{CH93}.

\subsection{Conclusions}

In conclusion,
we have presented an omnibus embedding methodology
for joint optimization of fidelity and commensurability
that allows us to address the manifold matching problem
by jointly identifying embeddings of multiple spaces into a common space.
Such a joint embedding facilitates statistical inference
in a wide array of disparate information fusion applications.
We have investigated this methodology in the context of simple statistical inference tasks,
and compared and contrasted with competing fidelity-only and commensurability-only
methodologies, demonstrating the superiority of our joint optimization.

We have focused on a simple setting and simple choices for various methodological options.
Many variations and generalizations are possible,
but the presentation here provides the core methodological instantiation.

\bibliographystyle{unabuser}
\bibliography{jofcR1}

\end{document}